\documentclass[prb,final,twocolumn,showpacs,showkeys]{revtex4}
\usepackage{amsmath}
\usepackage{graphicx}
\usepackage{amsfonts}
\usepackage{amssymb}
\usepackage{subfigure}
\begin{document}

\title{Perfect Quantum State Transfer with Superconducting Qubits}

\author{Frederick W. Strauch}
\email[Electronic address: ]{Frederick.W.Strauch@williams.edu}
\altaffiliation[Current address: ]{Williams College, Williamstown, MA 01267}
\affiliation{National Institute of Standards and Technology, Gaithersburg, Maryland 20899-8423, USA}
\author{Carl J. Williams}
\affiliation{National Institute of Standards and Technology, Gaithersburg, Maryland 20899-8423, USA}

\date{\today}

\begin{abstract}
Superconducting quantum circuits, fabricated with multiple layers, are proposed to implement perfect quantum state transfer between nodes of a hypercube network.  For tunable devices such as the phase qubit, each node can transmit quantum information to any other node at a constant rate independent of the distance between qubits.  The physical limits of quantum state transfer in this network are theoretically analyzed, including the effects of disorder, decoherence, and higher-order couplings.  
\end{abstract} 
\maketitle

\section{Introduction}

Quantum information processing requires a large number of highly interconnected qubits.  However, most leading experimental candidates suffer from limited connectivity.  Typical designs involve low-dimensional networks with nearest-neighbor interactions, such as spins of electrons \cite{Loss98} or nuclei \cite{Kane98} in solids, or atoms in optical lattices.\cite{Jaksch99}  Other designs have networks of a limited number of qubits interacting through a common mode, such as ion traps.\cite{Cirac95}  In all these cases, the underlying spatial arrangement leads to a minimum time for information to be routed between the most distant qubits, a challenge to scalability.  Moving quantum information through the computer by swap gates, or moving the qubits themselves, are plausible but potentially slow, error-ridden processes.  More sophisticated proposals include measurement-based teleportation protocols \cite{Brennen2003b} and coupling of distant qubits by photons; \cite{Cirac97} both pose additional experimental challenges.  Efficient quantum routing remains a significant design problem for quantum-computer architectures. 

Superconducting circuits are a remarkable exception, as their couplings are controlled by the fabrication of superconducting wires.  This wiring can be complex, involving multiple layers (using interconnects and crossovers), \cite{Satoh2005} and thus is capable of three- or higher-dimensional topologies.  From a quantum information perspective, this resource in connectivity can be exploited in novel architectures, \cite{Fowler2007} and can even be integrated with other physical qubits such as atoms, \cite{Sorenson2004} ions,\cite{Tian2004} or polar molecules. \cite{Rabl2006}  From a physics perspective, studying these new artificial solids opens up a number of opportunities.  

The simplest such study is quantum transport, particularly the coherent transfer of one qubit state to another.  As shown by Bose, state transfer in a one-dimensional chain, using time evolution of a fixed Hamiltonian, can efficiently distribute entanglement over large distances \cite{Bose2003} without photons or feedback.  Christandl {\it et al.} studied networks implementing perfect state transfer\cite{Christandl2004} and Feder has recently shown how to generate an infinite hierarchy of such networks.\cite{Feder2006}  One such network is the hypercube, well-known in classical computer design,\cite{Hayes89} and even the subject of investigation by Feynman one summer years ago. \cite{Hillis89}

\begin{figure}
\begin{center}
\includegraphics[width = 3 in]{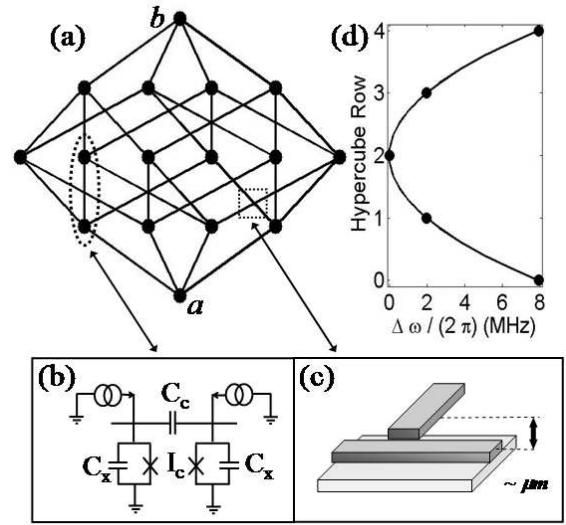}
\caption{(a) Hypercube network of dimension $d=4$, with corners labeled $a$ and $b$.  (b) Each node is implemented as a current biased Josephson junction, while each connection requires a coupling capacitor.  Typical circuit parameters include $C_x = 6$ pF, $C_c = 60$ fF, $I_c = 21 \mu$A, and $\omega_0 /2\pi = 6$ GHz.  (c) Crossing lines in the network are implemented by superconducting wires in a crossover configuration, using either insulating material or vacuum gaps between the layers.  For sufficiently large gaps ($100$ nm - $1\mu$m) the capacitance between the wires can be made negligible.  (d) A quadratic frequency shift of each row of qubits in the hypercube is required for high-fidelity state transfer (see text).}
\label{circuit}
\end{center}
\end{figure}

We propose to use multi-layered superconducting circuits of capacitively-coupled phase qubits \cite{Ramos2003} to implement hypercube quantum state transfer.  Coherent manipulations of single phase qubits using Rabi oscillations have been successful in a number of recent experiments,\cite{Martinis2002} while coupled systems have also been studied,\cite{Berkley2003} including full quantum state tomography of entanglement.\cite{Steffen2006b} In addition to direct coupling of two qubits, superconducting circuits allow for coupling through a third resonant circuit.  This was first demonstrated in spectroscopy experiments with phase qubits coupled to a lumped-circuit resonator \cite{Xu2005} and more recently in time-domain experiments involving the active emission and absorption of a photon in a transmission line coupling two phase qubits\cite{Simmonds2007} and through dispersive interactions of charge qubits with an intermediate cavity. \cite{Majer2007}  

While we focus on phase qubits coupled by capacitors in the following, our results can be readily extended to these and other superconducting qubit designs.   Previous theoretical studies of one-dimensional quantum state transfer have been performed for both charge \cite{Romito2005} and flux \cite{Lyakhov2005} qubits.   More generally, the fundamental importance of novel quantum transport in superconducting qubit arrays was discussed by Levitov {\it et al.}\cite{Levitov2001}  Central to our multi-dimensional circuit design is the use of long-range couplings and multi-layered devices---these properties have also been used in other recent designs for superconducting quantum computers. \cite{Fowler2007} 

In the following sections, we first describe the design of a phase qubit hypercube.  Then, by using the tunability of the phase qubit, we show how high fidelity state transfer can be performed between any two of the $2^d$ qubits in the $d$-dimensional hypercube network.  Furthermore, we show how to correct errors due to the higher-order coupling terms naturally present in the multi-qubit Hamiltonian.  Finally, we analyze the effects of decoherence and disordered couplings to show that this approach can be demonstrated using existing technology.  

\section{Phase Qubit Hypercube}

We consider a network of capacitively-coupled phase qubits \cite{Ramos2003} (see Fig.~\ref{circuit}), modeled as current-biased Josephson junctions described by the following Hamiltonian
\begin{equation}
\mathcal{H} = \frac{1}{2} \left(\frac{2 \pi}{\Phi_0}\right)^2 \sum_{j k} p_j (C^{-1})_{jk} p_k - \sum_{j} \frac{\Phi_0}{2\pi} \left(I_{cj} \cos \gamma_j + I_j \gamma_j\right),
\end{equation}
where junction $j$ has a critical current $I_{cj}$ and adjustable bias current $I_j$, and where $\Phi_0=h/(2e)$ is the flux quantum.  The dynamical variables are the phase differences of each junction $\gamma_j$ and their corresponding conjugate momenta $p_j$, the latter coupled by (the inverse of) the capacitance matrix $C_{jk}$.  In the ideal case (to be relaxed in Sec. V), all junctions have the same critical current $I_{cj} = I_c$, intrinsic capacitance $C_x$, and are coupled by identical capacitors $C_c$.  The capacitance matrix then takes the form $C_{jk} = C (\delta_{jk} - \zeta A_{jk}$), where $C = C_x + d C_c$ is the total capacitance of each junction, $\zeta = C_c / C = C_c/(C_x + d C_c)$ is a dimensionless coupling parameter, and $A_{jk}$ is the adjacency matrix for the $d$-dimensional hypercube (see below).

By expanding the Hamiltonian in terms of the lowest two eigenstates of each uncoupled phase qubit, and utilizing a rotating wave approximation, we find
\begin{equation}
\mathcal{H} \approx -\frac{1}{2} \sum_{j} \hbar \omega_j Z_j + \frac{1}{2} \sum_{j < k} \hbar \Omega_{j k} \left( X_j X_k + Y_j Y_k \right)
\label{qbit1}
\end{equation}
where $X$, $Y$, and $Z$ are the Pauli matrices for each qubit, $\omega_j \approx (2\pi I_c (C^{-1})_{jj}/ \Phi_0)^{1/2} (1 - I_j^2/I_c^2)^{1/4}$, and the coupling matrix is
\begin{equation}
\Omega_{jk} = \frac{1}{2} \sqrt{\omega_j \omega_k} \left( \zeta A + \zeta^2 A^2 + \zeta^3 A^3 + \cdots \right)_{jk}.
\label{qbit2}
\end{equation}

The adjacency matrix can be written most efficiently by first labeling each vertex by a binary number $x$ with $d$ bits, i.e. $x = x_1 x_2 \cdots x_d$ with each $x_j$ equal to $0$ or $1$.  A hypercube will result if we connect vertices whose labels differ only by one bit.  Thus, the corner labeled $x = 000...0$ will have $d$ neighbors, each with a single $1$ in its label (e.g. $y=100...0$).  The elements $A_{x,y}$ of the adjacency matrix are unity whenever the labels correspond to connected vertices, and are zero otherwise.  

Using the binary labeling one can introduce a tensor-product structure on the adjacency matrix.  Letting $\tau$ be the Pauli matrix $\tau = \sigma_x$ operating in this vertex space, one has
\begin{equation}
A = \sum_{j=1}^d \tau_j,
\label{adjmat}
\end{equation}
where each term $\tau_j$ in the sum corresponds to the tensor product of $\tau$ acting on the binary index $j$ with the identity matrix for every other index: 
\begin{equation}
(\tau_j)_{x,y} = \tau_{x_j,y_j} \prod_{k \ne j} \delta_{x_k,y_k}.
\end{equation}
For example, the adjacency matrix of the cube ($d=3$)
\begin{equation}
A = \left( \begin{array}{cccccccc}
0 & 1 & 1 & 0 & 1 & 0 & 0 & 0 \\
1 & 0 & 0 & 1 & 0 & 1 & 0 & 0 \\
1 & 0 & 0 & 1 & 0 & 0 & 1 & 0 \\
0 & 1 & 1 & 0 & 0 & 0 & 0 & 1 \\
1 & 0 & 0 & 0 & 0 & 1 & 1 & 0 \\
0 & 1 & 0 & 0 & 1 & 0 & 0 & 1 \\
0 & 0 & 1 & 0 & 1 & 0 & 0 & 1 \\
0 & 0 & 0 & 1 & 0 & 1 & 1 & 0 \end{array}\right)
\end{equation}
can be broken into the the sum of
\begin{equation}
\tau_1 = \left( \begin{array}{cccccccc}
0 & 0 & 0 & 0 & 1 & 0 & 0 & 0 \\
0 & 0 & 0 & 0 & 0 & 1 & 0 & 0 \\
0 & 0 & 0 & 0 & 0 & 0 & 1 & 0 \\
0 & 0 & 0 & 0 & 0 & 0 & 0 & 1 \\
1 & 0 & 0 & 0 & 0 & 0 & 0 & 0 \\
0 & 1 & 0 & 0 & 0 & 0 & 0 & 0 \\
0 & 0 & 1 & 0 & 0 & 0 & 0 & 0 \\
0 & 0 & 0 & 1 & 0 & 0 & 0 & 0 \end{array}\right),
\end{equation}
\begin{equation}
\tau_2 = \left( \begin{array}{cccccccc}
0 & 0 & 1 & 0 & 0 & 0 & 0 & 0 \\
0 & 0 & 0 & 1 & 0 & 0 & 0 & 0 \\
1 & 0 & 0 & 0 & 0 & 0 & 0 & 0 \\
0 & 1 & 0 & 0 & 0 & 0 & 0 & 0 \\
0 & 0 & 0 & 0 & 0 & 0 & 1 & 0 \\
0 & 0 & 0 & 0 & 0 & 0 & 0 & 1 \\
0 & 0 & 0 & 0 & 1 & 0 & 0 & 0 \\
0 & 0 & 0 & 0 & 0 & 1 & 0 & 0 \end{array}\right),
\end{equation}
and
\begin{equation}
\tau_3 = \left( \begin{array}{cccccccc}
0 & 1 & 0 & 0 & 0 & 0 & 0 & 0 \\
1 & 0 & 0 & 0 & 0 & 0 & 0 & 0 \\
0 & 0 & 0 & 1 & 0 & 0 & 0 & 0 \\
0 & 0 & 1 & 0 & 0 & 0 & 0 & 0 \\
0 & 0 & 0 & 0 & 0 & 1 & 0 & 0 \\
0 & 0 & 0 & 0 & 1 & 0 & 0 & 0 \\
0 & 0 & 0 & 0 & 0 & 0 & 0 & 1 \\
0 & 0 & 0 & 0 & 0 & 0 & 1 & 0 \end{array}\right).
\end{equation}

An experimental implementation of this design requires a fabrication process using multiple layers.  The hypercube topology can be accomplished with only pairwise crossing of the coupling wires.  These wires can be fabricated in a cross-over configuration, shown in Fig. 1(c), in which insulating material separates the two superconducting layers.  By making this separation large enough ($\sim 1 \mu$m), cross-coupling between these wires can be reduced significantly.  Multi-layered superconducting circuits with six metallic layers have been reported. \cite{Satoh2005}

In addition, coupling a large number of qubits will require wires spanning multiple qubits.  A long wire will have a non-negligible inductance $L$, leading to a resonant mode with frequency $\omega_{LC} = 1/\sqrt{L C_c}$ which could perturb the state transfer dynamics.  However, long-range coupling of this sort has already been tested \cite{Xu2005} with lengths approaching one millimeter, large compared to the typical dimensions of the qubits ($\sim 10\mu$m).  The corresponding inductance in this circuit was found to be $L \sim 2$ nH, which combined with a coupling capacitance $C_c = 30$ fF leads to $\omega_{LC}/2\pi > 20$ GHz, much higher than the qubit frequency ($\omega_0/2\pi \sim 6$ GHz).    This is significantly off-resonant and thus the resonant mode will remain in its ground state---although the nature of the coupling will be slightly modified. \cite{Xu2005}

Finally, the actual fabrication of a hypercube circuit appears quite complex.  However, the number of wires in this design is actually quite modest.  The hypercube circuit can couple any two qubits in the same amount of time---this is shown in the next section.  This property is shared by a completely connected set of $N$ qubits.  However, since a $d$-dimensional hypercube has $d \, 2^{d-1}$ edges, \cite{Hayes89} the hypercube requires significantly less connections.  Specifically, the completely connected circuit would require $N (N-1)/2$ wires ($\sim 10^6$ for $N = 1024$), while the hypercube would require only $(N/2) \log_2 N$ wires ($\sim 5000$ for $N=1024$).   We conclude that this is a promising theoretical design that can be implemented, at least for modest $d$, using existing technology.

\section{Perfect State Transfer}

Quantum state transfer involves the preparation of an initial state $|\Psi_i \rangle = \left(\alpha |0\rangle + \beta |1\rangle \right)_{a} \otimes |0\rangle_{\text{rest}} \otimes |0\rangle_{b}$, containing quantum information in the qubit $a$, and evolution by the time-independent Hamiltonian $\mathcal{H}$:
\begin{equation}
|\Psi(t)\rangle = e^{-i \mathcal{H} t/\hbar} \left(\alpha |0\rangle + \beta |1\rangle \right)_{a} \otimes |0\rangle_{\text{rest}} \otimes |0\rangle_{b}.
\label{sxf0}
\end{equation}
The quantum information will propagate through the network (the ``rest''), potentially to emerge in qubit $b$ at time $T$:
\begin{equation}
|\Psi(t=T)\rangle \approx |\Psi_f\rangle = |0\rangle_{a} \otimes |0\rangle_{\text{rest}} \otimes \left(\tilde{\alpha} |0\rangle + \tilde{\beta} |1\rangle\right)_{b}.
\label{sxf1}
\end{equation}

Following the work of Bose, \cite{Bose2003} the dynamics is of Eq.~(\ref{qbit1}) is restricted to the states $|0) = |0 \dots 0\rangle$ (the ground state) and the $N$ first-excited states $|x) = X_x |0)$, where $X_x$ is the Pauli operator for the qubit at site $x$.  Specifically, starting with the initial state $|\Psi(0)\rangle = \alpha |0) + \beta |a)$, the time evolution given by $\mathcal{H}$ yields the final state
\begin{equation}
|\Psi(t)\rangle = \alpha e^{i \phi} |0) + \beta \sum_{x} f_x(t) |x),
\end{equation}
with $\phi = \frac{1}{2} \sum_j \omega_j t $
and
\begin{equation}
f_x(t) = e^{i \phi} \left(\exp[-i (\omega + \Omega)t] \right)_{x,a}.
\end{equation}
Here $\omega$ is a diagonal matrix (with elements $\omega_j$) and $\Omega$ the coupling matrix given by Eq.~(\ref{qbit2}).  Perfect state transfer, as described above, will occur if $|f_b(T)| = 1$.  Note that this expression shows that for this type of state transfer we need only work with the effective Hamiltonian for this subspace, with the matrix form $\mathcal{H}_{eff} = \hbar (\omega + \Omega)$.  We now show, using the results of Christandl {\it et al.}, \cite{Christandl2004} how the hypercube network can implement perfect state transfer from corner to corner.  

Letting $a$ and $b$ be the (labels for the) qubits at any two corners of the hypercube, as in Fig. 1(a), we first set $\omega_j = \omega_0$ in the qubit Hamiltonian (\ref{qbit1}) and approximate (\ref{qbit2}) by $\Omega_{jk} \approx \zeta \omega_0 A_{jk}/2$.  Then, evaluating $f_x(t)$ by using the tensor-product form of the adjacency matrix given by Eq.~(\ref{adjmat}) we find 
\begin{equation}
f_b(t) = e^{i \phi} e^{-i \omega_0 t} (-i)^d [\sin ( \zeta \omega_0 t / 2)]^d.
\end{equation}
Ignoring the overall phase $\phi$, we find that perfect state transfer occurs at the transfer time $T = \pi / ( \zeta \omega_0)$ with $\tilde{\alpha} = \alpha$ and $\tilde{\beta} = e^{-i \omega_0 T} (-i)^d \beta$.  Note that for phase qubits, the initial state can be prepared and the final state can be verified using the state tomography techniques demonstrated by Steffen {\it et al.} \cite{Steffen2006b}  

By dynamically switching the qubit frequencies into and out of resonance, this state transfer protocol can be extended to {\it any pair} of qubits in the hypercube.  This is done by setting $\omega_j = \omega_k$ only if $(j,k)$ is an edge of the subcube of the hypercube with $a$ and $b$ as corners, and $|\omega_j - \omega_k| > \zeta \omega_{j,k}$ otherwise.  That is, this network can be ``programmed'' by tuning the qubit frequencies into resonance for subcubes of the hypercube.  As there are many disjoint subcubes, \cite{Hayes89} this network allows for the transfer of quantum information in parallel, all with similar transfer times.  This dynamical switching can be achieved by low-frequency control of the phase qubit's control circuit, \cite{Strauch2003} and was demonstrated in a recent experiment. \cite{Simmonds2007} 

This network can be used for many tasks in quantum information processing.  By coupling each node of this network to an additional set of ``memory'' qubits, this scheme allows for the parallel execution of multi-qubit gates between a large ($2^d$) number of qubits.  A more promising application (from the perspective of fault tolerance) builds on this idea to quickly distribute entanglement between multiple nodes that can then be purified locally, and refreshed during computation.  This entanglement can then be used for logic gates, error detection, or other teleportation-based tasks. \cite{Knill2005}

\section{Higher-Order Couplings}

There are many limitations to the success of this scheme, which we now analyze.  The qubit and rotating wave approximations have been used to obtain Eq.~(\ref{qbit1}).  The qubit approximation fails if higher energy states are excited.  This could occur, for example, during parallel operation of subcubes, but can be avoided by sufficiently large detuning. \cite{Strauch2003}  The rotating wave approximation should hold for sufficiently small couplings $\zeta < 0.1$. \cite{Pritchett2005}  

A more significant error is due to the higher-order terms in the coupling matrix $\Omega$ in Eq.~(\ref{qbit2}).  State transfer succeeds by engineering the spectrum of the first-excited-state subspace.  This spectrum is determined by the adjacency matrix, which for  the hypercube can be mapped to an angular momentum operator on the vertex space: $A = 2J_x$.\cite{Christandl2004}  Using this mapping for the phase-qubit hypercube, the effective Hamiltonian $\mathcal{H}_{eff} = \hbar ( \omega + \Omega)$ can be written as
\begin{equation}
\mathcal{H}_{eff} = \hbar \omega_0 (\zeta J_x + 2 \zeta^2 J_x^2 + \cdots),
\end{equation} 
where we have ignored an overall constant ($\hbar \omega_0$).  While $J_x$ has a linear spectrum (leading to perfect transfer), the higher-order terms add quadratic (and higher) terms to the spectrum.  These terms lead to an intrinsic dephasing of the time evolution.  

However, by varying the frequencies of each qubit, one can correct for these terms by the following scheme.  First, label the rows of the $d$-dimensional hypercube from $0$ to $d$ (these labels are equal to the Hamming distance of the binary label of each node from $x_a = 0 \cdots 0$).  Then, setting the frequencies for each qubit in row $k$ to $\omega_0 + \omega_0 \beta_2 (k-d/2)^2$ leads to the new effective Hamiltonian: 
\begin{equation}
\mathcal{H}_{eff} \approx \hbar \omega_0 \left(\beta_2 J_z^2 + \zeta J_x + 2 \zeta^2 J_x^2 + \cdots \right). 
\end{equation}  
One can then show, using perturbation theory in the eigenstates of $J_x$, that the choice $\beta_2 = 4 \zeta^2$ corrects for the contribution of $J_x^2$ to the spectrum.  Thus, by simply engineering a quadratic frequency shift across the network, this lowest order error can be eliminated [see Fig. 1(d)].  

\begin{figure}
\begin{center}
\includegraphics[width = 3.5 in]{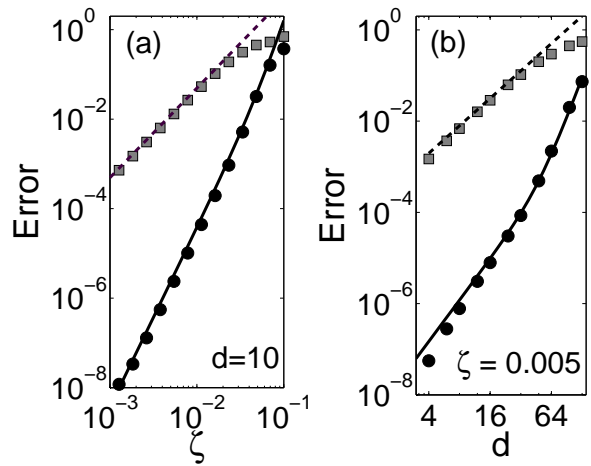}
\caption{Error in state transfer due to higher-order couplings.  (a) The error $1-F$ as a function of coupling strength $\zeta$, where $F$ is the state transfer fidelity for a $d=10$ hypercube with (circles) and without (squares) the quadratic frequency shift.  The upper (dashed) line is the function $\pi^2 d^2 \zeta^2/2$, while the lower (solid) line is $3 \pi^2 d^3 \zeta^4 /8 + \pi^2 d^6 \zeta^6/8$; these are found from a perturbative solution.  (b) The error $1-F$ as a function of hypercube dimension $d$ with $\zeta = 0.005$.  }
\label{errorx}
\end{center}
\end{figure}

We now demonstrate the effectiveness of this scheme by calculating how the error $1-F$ scales as a function of $\zeta$ and $d$, shown in Fig. \ref{errorx}.  Here $F = |\langle \Psi_f | e^{-i \mathcal{H}_{eff} T/\hbar} | \Psi_i\rangle|^2$ is the fidelity calculated using the effective Hamiltonian, simulating state transfer with the initial state specified by $\alpha = 0$ and $\beta = 1$.  Without any correction, the error scales as $d^2 \zeta^2$, shown by the dashed lines, while the correction described above yields residual errors proportional to $d^3 \zeta^4$ and $d^6 \zeta^6$, shown by the solid lines.  These curves are found by calculating the effect of the quadratic and cubic terms of the eigenvalue spectrum on the state transfer fidelity.  These terms arise from both the higher-order couplings omitted above (e.g. $\zeta^3 A^3$) and the imposed variation of the qubit frequencies.  As these qubit frequencies can also be controlled dynamically, these higher-order errors could be eliminated by using well-known quantum control techniques. \cite{Vandersypen2004}  Nevertheless, this single correction reduces the error over two orders of magnitude, and becomes tolerable even for large ($d \sim 10-20$) hypercube networks.

\section{Decoherence}

Any experimental demonstration will also encounter the effects of decoherence and disorder.  We model the first by a master equation for the density matrix:
\begin{equation}
\begin{array}{ll}
\partial_t \rho = & -i [\mathcal{H}/\hbar,\rho] + \sum_{j} T_1^{-1} \left(\sigma^{-}_j \rho \sigma^{+}_j - \frac{1}{2} \{ \sigma^{+}_j \sigma^{-}_j, \rho \} \right) \\
& + \frac{1}{2} \sum_{j} T_{\phi}^{-1} \left(Z_j \rho Z_j - \rho \right),
\end{array}
\label{dmat1}
\end{equation}
with $\sigma^{\pm}_j =(X_j \mp i Y_j)/2$.  Here we have introduced independent energy decay ($T_1$, also called amplitude damping) and dephasing ($T_{\phi}$, also called phase damping) processes for each qubit. \cite{Vandersypen2004}  For weak decoherence [$T_1, T_{\phi} \gg T = \pi/(\zeta \omega_0)$] we now perturbatively solve this master equation given the initial condition $\rho(0) = |\Psi_i\rangle \langle \Psi_i|$.  

We first note that, given the initial condition, $\rho(t)$ can be written as 
\begin{equation}
\begin{array}{lcl}
\rho(t) &=& \rho_{0,0}(t) |0) (0| + \sum_{x} \left[ \rho_{0,x} (t) |0) (x| + \rho_{x,0}(t) |x)(0| \right] \\
& & + \sum_{x,y}\rho_{x,y}(t) |x)(y|.
\end{array}
\end{equation}
Note that $\rho_{0,0}(0) = |\alpha|^2$, $\rho_{0,x}(0) = \alpha \beta^* \delta_{a,x}$, $\rho_{x,0}(0) = \alpha^* \beta \delta_{a,x}$, and $\rho_{x,x}(0) = |\beta|^2 \delta_{a,x}$.  The density matrix elements satisfy the differential equations:
\begin{equation}
\begin{array}{lcl}
\dot{\rho}_{0,0} &=& T_1^{-1} \sum_{x} \rho_{x,x} = T_1^{-1} (1-\rho_{0,0}) \\
\dot{\rho}_{0,x} &=& i \omega_{x} \rho_{0,x} + i \sum_{y} \Omega_{x,y} \rho_{0,y} - T_2^{-1} \rho_{0,x} \\
\dot{\rho}_{x,0} &=& -i \omega_{x} \rho_{x,0} - i \sum_{y} \Omega_{x,y} \rho_{y,0} - T_2^{-1} \rho_{x,0} \\
\dot{\rho}_{x,y} &=& -i \sum_{z} ( \Omega_{x,z} \rho_{z,y} - \rho_{x,z} \Omega_{z,y}) \\
& &  - T_1^{-1} \delta_{x,y} \rho_{x,x} - 2 T_2^{-1} \rho_{x,y}(1-\delta_{x,y}).
\end{array}
\label{masteq}
\end{equation} 

The first equation in (\ref{masteq}) can be solved directly to yield
\begin{equation}
\rho_{0,0}(t) = 1 - |\beta|^2 e^{-t/T_1}.
\end{equation}
Using the conditions $\omega_x = \omega_0$ and $ \Omega_{x,y} \approx \zeta \omega_0 A_{x,y}/2$ as before, we can use the results of Section III to solve the second and third equations in (\ref{masteq}) (with $x=b$)
\begin{equation}
\begin{array}{lcl}
\rho_{0,b}(t) &=& \alpha \beta^* e^{-t/T_2} e^{+i \omega_0 t} (+i)^d [ \sin (\zeta \omega_0 t/2)]^d \\
\rho_{b,0}(t) &=& \alpha^* \beta e^{-t/T_2} e^{-i \omega_0 t} (-i)^d [ \sin (\zeta \omega_0 t/2)]^d,
\end{array}
\end{equation}
where $T_2^{-1} = (2 T_1)^{-1} + T_{\phi}^{-1}$.

\begin{figure}
\begin{center}
\includegraphics[width = 3.5 in]{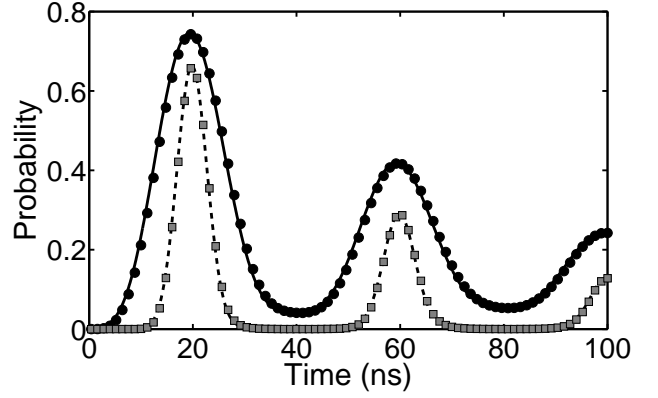}
\caption{Probability $\rho_{b,b}(t)$ as a function of time for state transfer with $\beta = 1$.  The symbols correspond to numerical simulations of the master equation with $\omega_0/2\pi = 5$ GHz, $\zeta = 0.005$, $T_1 = T_{\phi} = 120$ ns, for hypercubes of dimension $d=2$ (circles) and $d=10$ (squares).  The corresponding lines are evaluations of the perturbative solution given by Eq. (\ref{rhobb}).}
\label{rhofig}
\end{center}
\end{figure}

The equation for $\rho_{b,b}$, however, cannot be solved exactly.  A perturbative solution can be found, but requires a careful examination of the eigenstates of $\Omega$ and how they are mixed by the decoherence terms.  After a long calculation \cite{Strauchx} we find
\begin{equation}
\rho_{b,b}(t) = |\beta|^2 e^{-t/T_1} \sum_{n=0}^d (-1)^{d-n} g_n(t) \cos ( \zeta \omega_0 t (d-n)),
\label{rhobb}
\end{equation}
where  we have defined the functions
\begin{equation}
g_n(t) = \sum_{p=0}^{\lfloor n/2 \rfloor}\frac{d! (2 - \delta_{n,d}) 2^{n-2d -2p} }{p! (n-2p)! (d-n+p)!}  e^{- \lambda_{pn} t}
\end{equation}
and the decay constants
\begin{equation}
\lambda_{pn} = \frac{2}{T_{\phi}} \left(1 - 2^{n-d-2p} \frac{(d-n+2p)!}{p! (d-n+p)!}\right).
\end{equation}
As shown in Fig. \ref{rhofig}, this solution for $\rho_{b,b}(t)$ compares well to direct numerical simulations of the master equation (\ref{dmat1}) as well as to more detailed simulations (not shown) of a master equation for the phase qubit system involving a large number of non-qubit states (e.g. hundreds of states for $d=3$).   

Using these results we can calculate the state transfer fidelity $F = \langle \Psi_f | \rho(T) |\Psi_f\rangle$, at the transfer time $T = \pi/(\zeta \omega_0)$ with $|\Psi_f\rangle = \alpha |0) + (-i)^d e^{-i\omega_0 T} \beta |b)$:
\begin{equation}
\begin{array}{lcl}
F &=& |\alpha|^2 - |\beta|^2 e^{-T/T_1} + 2 |\alpha|^2 |\beta|^2 e^{-T/T_2} \\
& & + |\beta|^4 e^{-T/T_1} \sum_{n=0}^d g_n(T).
\end{array}
\end{equation}
An important feature of this solution is that, so long as the frequency $\omega_0$ and the coupling $\zeta$ remain fixed, the state transfer fidelity has a lower bound that is independent of the size of the network.  To show this, we follow Bose \cite{Bose2003} and consider the fidelity $F_{avg}$ averaged over the Bloch sphere (the worst case can be bounded similarly):   
\begin{equation}
F_{avg} = \frac{1}{2} - \frac{1}{6} e^{-T/T_1} + \frac{1}{3} e^{-T/T_2} + \frac{1}{3} e^{-T/T_1} \sum_{n=0}^{d} g_n(T).
\end{equation}
Using the fact that $\lambda_{pn} \le 2/T_{\phi}$, we find that $\sum_{n=0}^{d} g_n(T) \ge e^{-2 T/T_{\phi}}$ and thus
\begin{equation}
F_{avg} \ge \frac{1}{2} - \frac{1}{6} e^{-T/T_1} + \frac{1}{3} e^{-T/T_2} + \frac{1}{3} e^{-2 T/T_2}.
\end{equation}
Since the transfer time $T = \pi / (\zeta \omega_0)$ is independent of $d$, so is this bound on the fidelity.  Furthermore, this simple formula shows that the fidelity can be greater than the classical limit \cite{Bose2003} of $2/3$ for existing experimental devices.  For example, with $\omega_0/2\pi = 5$ GHz, $\zeta = 0.005$, and the decoherence times $T_1 = 120$ ns and $T_2 = 80$ ns, \cite{Steffen2006b} we find $T = 20$ ns and thus $F_{avg} \ge 0.8$.  

Note that this bound on the fidelity is specific to our decoherence model.  If the qubit decoherence times depend on the size of the circuit, this model should be improved.  Physically, the complexity of the wiring design (the cross-overs and interconnects) could introduce significant loss using traditional dielectrics. \cite{Martinis2005}  However, advanced fabrication using vacuum gaps should remove this potential decoherence source. \cite{Simmonds2007b}  Note also that for fixed capacitors (instead of fixed $\omega_0$ and $\zeta$),  there is in fact a weak dependence of the transfer time on $d$: $T(d) \approx T(1) [1+(d-1) \zeta]^{3/2}$.  However, for the experimental parameters given above, this dependence can be neglected for $d<20$.

\section{Disordered Couplings}

No experimental circuit will have identical qubits---there will always be some variation.  This variation in parameters is equivalent to disorder, and in this context we might expect a blocking of transport due to Anderson localization. \cite{Lee1985}  That is, by mapping propagation through the hypercube to an effective spin, the information travels along a one-dimensional path (row by row). \cite{Christandl2004}  This was studied in previous work, \cite{DeChiara2005} in which static disorder was introduced into the one-dimensional effective Hamiltonian for perfect state transfer.  There it was found that disordered couplings play a much larger role than disordered qubit frequencies, with results in agreement with localization theory.  Note that we have already allowed the qubit frequencies to be under experimental control, so here we only consider disordered couplings.

\begin{figure}
\begin{center}
\includegraphics[width = 3.5 in]{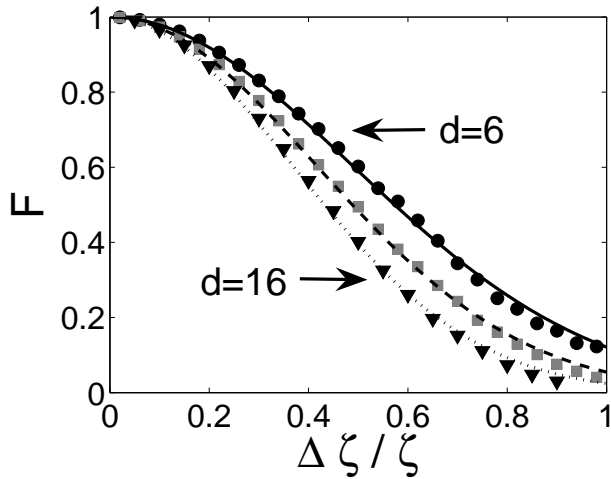}
\caption{Fidelity of state transfer due to disordered couplings.  Each point for $d=6$ (circles), $d=10$ (squares), and $d=16$ (triangles) represents the fidelity averaged over many (10000, 1000, and 100, respectively) realizations of disordered couplings (see text).  The corresponding lines (solid, dashed, and dotted) are fitted Gaussians.}
\label{errory}
\end{center}
\end{figure} 

To simulate the effects of static disorder on multi-dimensional state transfer, we use use the full subspace Hamiltonian and introduce disorder into the coupling matrix by $\Omega_{jk} = \omega_0 (\zeta + z_{jk}) A_{jk}/2$ (the higher-order terms are neglected here) where each $z_{jk}$ is a random variable uniformly distributed over the range $-\Delta \zeta < z_{jk} < \Delta \zeta$.  The transfer fidelity (with $\alpha=0$ and $\beta=1$) is then calculated and averaged over this uniform distribution.  The averaged fidelity is shown in Fig. \ref{errory}.  These results are consistent with $F \sim e^{-d/\ell}$, where $\ell \sim c (\Delta \zeta)^{-2}$.  This implies that, for very large $d$, transport is exponentially suppressed due to localization.  However, even a hypercube of modest dimension, where localization effects are small, can accomodate a large ($2^d$) number of qubits.  For example, while state-of-the-art fabrication can achieve variation of order 1\% or less, Fig. \ref{errory} shows that even for 10\% variation in a network with $d=10$ one still has $F > 0.95$.  

\section{Conclusion}

In conclusion, we have proposed that the hypercube network can be used to route quantum information between superconducting qubits.  Using tunable phase qubits, this network can be programmed to achieve high-fidelity quantum state transfer between any two nodes and allows for rapid distribution of entanglement in a superconducting quantum computer.  The dominant error mechanisms have been analyzed, including higher-order couplings, decoherence, and disorder.  Our results indicate that using existing qubit fabrication and experimental methods, a demonstration of these operations can be performed.  More generally, these initial results motivate continued study of novel quantum transport in these artificial solids.

\begin{acknowledgments}
We thank R. Simmonds,  R. Ramos, and D. Schuster for helpful comments and discussions.
\end{acknowledgments}

\end{document}